\documentstyle[twoside,fleqn,espcrc2,epsfig]{article}
\title{                         
Solutions to the Atmospheric Neutrino Problem}
\author{
M.\ C.\ Gonzalez-Garcia
\thanks{
Talk given at the Xth Baksan School "Particles and Cosmology", 
19-25 April 1999, Baksan Valley, Kabardino-Balkaria, Russia, 
and at the 5th International Workshop 
``Valencia 99: Particles in Astrophysics and Cosmology'', 
Valencia May 3-8, 1999.}
\address{Instituto de F\'{\i}sica Corpuscular -- C.S.I.C. \\
    Departamento de F\'{\i}sica Te\`orica, Universitat of Val\`encia \\
    46100 Burjassot, Val\`encia, Spain\\}}
\begin{document}
\begin{abstract}
In this talk I review the present status of the atmospheric neutrino anomaly
and discuss some solutions that have been presented in the literature
to solve this problem. In particular I review the "standard" solution
in terms of neutrino oscillations as well as alternative scenarios
such as the possibility of flavour changing neutrino interactions with
the Earth and neutrino decay.
\end{abstract}
\maketitle

\section{Introduction}
Neutrinos produced as decay products in hadronic showers from cosmic
ray collisions with nuclei in the upper atmosphere have
been observed by several detectors
\cite{Super-Kamiokande,sk99,Frejus,Nusex,Kamiokande,IMB,Soudan}.  Although
the absolute fluxes of atmospheric neutrinos are largely uncertain,
the expected ratio $(\mu/e)$ of the muon neutrino flux ($\nu_\mu +
\bar{\nu}_\mu$) over the electron neutrino flux ($\nu_e+\bar{\nu}_e$)
is robust, since it largely cancels out the uncertainties associated
with the absolute flux.  In fact, this ratio has been calculated
~\cite{fluxes} with an uncertainty of less than 5\% over energies
varying from 0.1~GeV to 100~GeV. On this resides our confidence in the
long-standing atmospheric neutrino anomaly.

Super-Kamiokande high statistics
observations~\cite{Super-Kamiokande} indicate that the deficit in the
total ratio R($\mu/e$) is due to the number of neutrinos arriving in
the detector at large zenith angles. The  $e$-like events do not
present any compelling evidence of a zenith-angle dependent suppression 
while the $\mu$-like event rates are substantially suppressed at large zenith
angles. The $\nu_\mu \to \nu_\tau$ as well as the $\nu_\mu \to \nu_s$ 
\cite{atm98,atmo98} oscillation hypothesis provides a very good 
explanation for this smaller-than-expected ratio, which is also simple
and well-motivated theoretically. This led the Super-Kamiokande
Collaboration to conclude that their data provide good evidence for
neutrino oscillations and neutrino masses \cite{skos}.  However,
alternative explanations to the atmospheric neutrino data have been
proposed in the literature including the possibility of 
flavour changing (FC) neutrino interactions in matter \cite{fcnc} and 
neutrino decay \cite{decay}, the violation of relativity principles~
\cite{vep,vli} or the violation of CPT symmetry \cite{vcpt}.  

In this talk I will review the "standard" solution
in terms of neutrino oscillations as well as alternative scenarios
such as the possibility of flavour changing neutrino interactions with
the Earth and neutrino decay.

\section{Atmospheric Neutrino Induced Events at Underground Experiments}

Atmospheric neutrinos can be detected in underground detectors by direct 
observation of their charged current interaction inside the detector.
These are the so called contained events. Contained events can be further 
classified into fully contained events when the produced charged lepton 
(either electron or muon) in  the neutrino interaction does not escape the 
detector, and partially contained muons when the produced muon exits the 
detector. Super-Kamiokande has divided their contained data sample into
sub-GeV events with visible energy below 1.2 GeV and multi-GeV above such
cutoff. On average, sub-GeV events arise from neutrinos of several hundreds of
MeV while multi-GeV events are originated by neutrinos with energies of the
order of several GeV. Higher energy muon neutrinos
and antineutrinos can also be detected indirectly by observing the muons
produced in their charged current interactions in the vicinity of the
detector. These are the so called upgoing muons. Should the muon 
stop inside the detector, it will be classified as a ``stopping'' muon,
(which arises from neutrinos of energies around ten GeV)
while if the muon track crosses the full detector the event is 
classified as a ``through-going'' muon which is originated by neutrinos
with energies of the order of hundred GeV..

Given certain neutrino conversion mechanism, the expected number of 
$\mu$-like and $e$-like contained events, $N_\alpha$, $\alpha = \mu, e$ 
can be computed as:
\begin{equation}
N_\mu= N_{\mu\mu} +\
 N_{e\mu} \; ,  \;\;\;\;\;\
N_e= N_{ee} +  N_{\mu e} \; ,
\label{eventsnumber}
\end{equation}
where
\begin{eqnarray}
N_{\alpha\beta} &=& n_t T
\int
\frac{d^2\Phi_\alpha}{dE_\nu d(\cos\theta_\nu)} 
\kappa_\alpha(h,\cos\theta_\nu,E_\nu) \nonumber \\
& & P_{\alpha\beta} \frac{d\sigma}{dE_\beta}\varepsilon(E_\beta)
dE_\nu dE_\beta d(\cos\theta_\nu)dh\;
\label{event0}
\end{eqnarray}
and $P_{\alpha\beta}$ is the conversion probability of $\nu_\alpha \to
\nu_\beta$ for given values of $E_{\nu}, \cos\theta_\nu$ and $h$,
i.e., $P_{\alpha\beta} \equiv P(\nu_\alpha \to \nu_\beta; E_\nu,
\cos\theta_\nu, h ) $.  In the Standard Model (SM), the only
non-zero elements are the diagonal ones, i.e. $P_{\alpha\alpha}=1$ for
all $\alpha$.
Here $n_t$ is the number of targets, $T$ is the experiment's running
time, $E_\nu$ is the neutrino energy and $\Phi_\alpha$ is the flux of
atmospheric neutrinos of type $\alpha=\mu ,e$; $E_\beta$ is the final
charged lepton energy and $\varepsilon(E_\beta)$ is the detection
efficiency for such charged lepton; $\sigma$ is the neutrino-nucleon
interaction cross section, and $\theta_\nu$ is
the angle between the vertical direction and the incoming neutrinos
($\cos\theta_\nu$=1 corresponds to the down-coming neutrinos).  In
Eq.~(\ref{event0}), $h$ is the slant distance from the production
point to the sea level for $\alpha$-type neutrinos with energy $E_\nu$
and zenith angle $\theta_\nu$. Finally, $\kappa_\alpha$ is the slant
distance distribution which is normalized to one \cite{pathlenght}.

The neutrino fluxes, in particular in the sub-GeV range, depend on the
solar activity.  In order to take this fact into account we use in
Eq.~(\ref{event0}) a linear combination of atmospheric neutrino fluxes
$\Phi_\alpha^{max}$ and $\Phi_\alpha^{min}$ which correspond to the
most active Sun (solar maximum) and quiet Sun (solar minimum),
respectively, with different weights, depending on the running period
of each experiment \cite{atm98}.

Upgoing muon data are usually presented in the form of measured muon fluxes.
We obtain the effective muon fluxes for both stopping and through-going
muons by convoluting the probabilities with the corresponding muon
fluxes produced by the neutrino interactions with the Earth. We
include the muon energy loss during propagation both in the rock and
in the detector according to ~\cite{muloss,ricardo} and we take into
account also the effective detector area for both types of events,
stopping and through-going.  
Schematically
\begin{equation}
\Phi_\mu(\theta)_{S,T}=\frac{1}{A(L,\theta)}\int 
\frac{d\Phi_\mu(E_\mu,\theta)}{dE_\mu} 
A_{S,T}(E_\mu,\theta)
\end{equation}  
where
\begin{eqnarray}
\frac{d\Phi_\mu}{dE_\mu} 
&=& \int\frac{d\Phi_{\nu_\mu}(E_\nu,\theta)}{dE_\nu} 
P_{\mu\mu} \frac{d\sigma}{dE_{\mu 0}} 
R(E_{\mu 0}, E_\mu) \nonumber \\ 
& & \kappa_\mu(h,\cos\theta_\nu,E_\nu) dE_{\mu 0} dE_\nu dh 
\end{eqnarray}
where $R(E_{\mu 0}, E_\mu)$ is the muon range function which accounts
for the muon energy loss during propagation. 
$A(L,\theta)=A_{S}(E_\mu,\theta)+A_{T}(E_\mu,\theta)$ 
is the projected detector area for internal pathlengths longer than $L$.
$A_{S}$ and $A_{T}$ are the corresponding areas for stopping 
and through-going muon trajectories.
For Super-Kamiokande we compute these effective 
areas using the simple geometrical picture given in Ref.~\cite{lipari1}. 

Following Ref.~\cite{atm98} we explicitly verify in our present
reanalysis the agreement of our predictions with the experimental
Monte Carlo predictions, leading to a good confidence in the
reliability of our results. 

\section{Conversion Probabilities}

For definiteness I assume a two-flavour scenario. For the oscillation
case one must solve the Schr\"oedinger evolution equation of the 
$\nu_\mu -\nu_X$ 
(where $X=e,\tau $ or $s$ sterile) system in the matter background for
{\sl neutrinos } 
\begin{eqnarray}
i{\mbox{d} \over \mbox{d}t}\left(\matrix{
\nu_\mu \cr\ \nu_X\cr }\right) & = & 
 \left(\matrix{
 {H}_{\mu}
& {H}_{\mu X} \cr
 {H}_{\mu X} 
& {H}_X \cr}
\right)
\left(\matrix{
\nu_\mu \cr\ \nu_X \cr}\right) \,\,,  \label{evolution1}\\
  H_\mu & \! = &  \! 
 V_\mu + \frac{\Delta m^2}{4E_\nu} \cos2 \theta_{\mu X} \\
 H_X & \!= 
& V_X -  \frac{\Delta m^2}{4E_\nu} \cos2 \theta_{\mu X},  \\
H_{\mu X}& \!= &  - \frac{\Delta m^2}{4E_\nu} \sin2 \theta_{\mu X},
\end{eqnarray}
where 
\begin{eqnarray}
\label{potential}
V_\tau=V_\mu & = &\frac{\sqrt{2}G_F \rho}{M} (-\frac{1}{2}Y_n)\,, 
\\
V_s &= & 0\,,\\
V_e & = & \frac{\sqrt{2}G_F \rho}{M} ( Y_e - \frac{1}{2}Y_n)
\\
\nonumber
\end{eqnarray}
Here $G_F$ is the Fermi constant, $\rho$ is the matter density in the
Earth, $M$ is the nucleon mass, and $Y_e$ ($Y_n$) is the electron
(neutron) fraction. I define $\Delta m^2=m_2^2-m_1^2$ in such a way
that if $\Delta m^2>0 \: (\Delta m^2<0)$ the neutrino with largest
muon-like component is heavier (lighter) than the one with largest
X-like component. For anti-neutrinos the signs of potentials $V_X$
should be reversed. In our calculations we have used the approximate 
analytic expression for the matter density profile in the Earth obtained in
Ref. \cite{lisi}. In order to obtain the oscillation probabilities
$P_{\alpha\beta}$ we have made a numerical integration of the
evolution equation. The probabilities for neutrinos and anti-neutrinos
are different because the reversal of sign of matter potential. Notice
that for the $\nu_\mu\to\nu_\tau$ case there is no matter effect and the
probability takes the well known form
\begin{equation}
P_{\mu\mu}=1-\sin^22\theta\sin^2(\frac{\Delta m^2 L}{2E}). 
\label{probosc}
\end{equation}
For the $\nu_\mu\to\nu_s$ case there are two possibilities depending on
the sign of $\Delta m^2$.  For $\Delta m^2 > 0$ the matter effects
enhance {\sl neutrino} oscillations while depress {\sl anti-neutrino}
oscillations, whereas for the other sign ($\Delta m^2<0$) the opposite
holds. In what follows I will not consider the possibility of
oscillation into electron neutrinos as it is known to be ruled
out by the negative results of the reactor experiment Chooz \cite{chooz}.

I am also going to consider the possibility of 
FC-neutrino interactions of massless neutrinos which can also induce 
$\nu_\mu \to \nu_\tau$ transitions \cite{fcnc}. 
In our phenomenological approach we have assumed that the evolution
equations which describe $\nu_\mu \to \nu_\tau$ transitions in matter 
may be written as Eq.(\ref{evolution1}) with  
\begin{eqnarray} 
  H_\mu = 0  &  & 
  H_\tau = \sqrt{2}\ G_F  \epsilon_{\nu} ' n_f(r) \\
 &  &  H_{\mu\tau} = \sqrt{2}\ G_F  \epsilon_{\nu}n_f(r) 
\end{eqnarray}
where $\sqrt{2}\,G_F n_f(r) \epsilon_\nu$ is the
$\nu_\mu+ f \to \nu_\tau + f$ forward scattering amplitude and
$\sqrt{2}\,G_F n_f(r) \epsilon_\nu '$ is the difference between the
$\nu_\tau - f$ and $\nu_\mu - f$ elastic forward scattering
amplitudes, with $n_f(r)$ being the number density of the fermions
which induce such processes. 
The parameters $\epsilon_\nu$ and $\epsilon'_\nu$ contain the information
about FC--neutrino interactions, for details I refer to Ref. \cite{fcnc}.
In order to obtain the oscillation probabilities
$P_{\alpha\beta}$ we have made a numerical integration of the
evolution equation. For the sake of illustration I show here
the solution in the approximation of constant
matter density. The conversion probability in this case is
\begin{equation}
P_{\mu\mu}= 1-
\frac{4\epsilon_\nu^2}{4\epsilon_\nu^2+{\epsilon_\nu^\prime}^2}
\sin^2({1\over 2} \eta L), 
\label{probfc}
\end{equation}
where $\eta = 
\sqrt{4\epsilon_\nu^2+{\epsilon^\prime_\nu}^2} \sqrt{2} G_F n_f$. 

Finally when discussing the possibility of neutrino decay in the
$\nu_\mu$ $\nu_\tau$ system one must include also the effect
of the neutrino instability when solving the time evolution of the
mass eigenstates. In this case the evolution equation can be solved
analytically and the muon neutrino survival probability is given by:
\begin{equation}
P_{\mu\mu}=
\sin^4\theta+\cos^4\theta  \exp\left(-\frac{m_2}{E}\frac{L}{\tau}\right)
+ P_{\Delta m^2} 
\label{probdec}
\end{equation}
with
\begin{displaymath}
P_{\Delta m^2}=\frac{1}{2}\sin^22\theta 
\cos\left(\frac{\Delta m^2 L}{2E} \right)
\exp\left(-\frac{m_2}{E}\frac{L}{2\tau}\right)
\end{displaymath}
where $\tau$ is the neutrino lifetime. For clarity, I 
isolated the term $P_{\Delta m^2}$, explicitly dependent on the mass 
difference, $\Delta m^2=m_2^2-m_1^2$. In the following I will neglect
this term which averages out in the limit 
$ m_1 \ll m_2 \ll E$. The factor $m_2/E$ in the exponential is just 
the $\gamma$ Lorentz factor. 

By comparing Eqs.(\ref{probosc}), (\ref{probfc}), and (\ref{probdec}) 
one can see that the different mechanisms lead to different energy
dependence of the survival probability. While oscillations give a
muon survival probability slowly growing with the energy, neutrino
decay leads to a much faster growth. On the other hand, flavour changing
neutrino interactions lead to an energy independent survival probability.
As pointed out in the previous section, data on contained events
and upgoing muons scan a range of neutrino energies over three orders
of magnitude from hundreds of MeV to hundreds of GeV. This is
of great importance when discriminating between the different 
scenarios as pointed out in Refs.\cite{lipari,lisiup}.

\section{Atmospheric Neutrino Data Fits} 

Here I describe our fitting method to determine the atmospheric
neutrino conversion parameters for the various possible conversion channels.
In doing so we have relied on the separate use of the event numbers paying 
attention to the correlations between the errors in the muon predictions 
and electron predictions as well as the correlations among the errors of
the different energy data samples. 

The steps required in order to generate the allowed regions of
oscillation parameters were given in Ref. \cite{atm98}.  
Following Ref. \cite{atm98,fogli2} we defined the $\chi^2$ as
\begin{equation}
\chi^2 \equiv \sum_{I,J}
(N_I^{da}-N_I^{th}) \cdot 
(\sigma_{da}^2 + \sigma_{th}^2 )_{IJ}^{-1}\cdot 
(N_J^{da}-N_J^{th}),
\label{chi2}
\end{equation}
where $I$ and $J$ stand for any combination of the experimental data
set and event-type considered, i.e, $I = (A, \alpha)$ and $J = (B,
\beta)$ where, $A,B$ stands for the different experiments or different
data samples in a given experiment.
$\alpha, \beta = e,\mu$.  In Eq.~(\ref{chi2}) $N_I^{th}$ is the
predicted number of events calculated from Eq.~(\ref{eventsnumber})
whereas $N_I^{da}$ is the number of observed events.  In
Eq.~(\ref{chi2}) $\sigma_{da}^2$ and $\sigma_{th}^2$ are the
error matrices containing the experimental and theoretical errors
respectively. They can be written as
\begin{equation}
\sigma_{IJ}^2 \equiv \sigma_\alpha(A)\, \rho_{\alpha \beta} (A,B)\,
\sigma_\beta(B),
\end{equation}
where $\rho_{\alpha \beta} (A,B)$ stands for the correlation between
the $\alpha$-like events in the $A$-type experiment and $\beta$-like
events in $B$-type experiment, whereas $\sigma_\alpha(A)$ and
$\sigma_\beta(B)$ are the errors for the number of $\alpha$ and
$\beta$-like events in $A$ and $B$ experiments, respectively. The
dimension of the error matrix varies depending on the combination of
experiments included in the analysis.  

We have computed $\rho_{\alpha \beta} (A,B)$ as in Ref. \cite{fogli2}.  A
detailed discussion of the errors and correlations used in our
analysis for the contained events can be found in Ref.~\cite{atm98}.  
In our present analysis, we have also included the data on stopping and
through-going muons. We have conservatively ascribed a 20\% uncertainty to the
absolute neutrino flux, in order to generously account for the spread
of predictions in different neutrino flux calculations. Other important
source of theoretical uncertainty arises from the neutrino interaction
cross section which at Super-Kamiokande ranges from 10--15 \%. 
We allow a 5\% variation in the ratio 
between muon events in different energy samples. We further introduce a
10\%  theoretical error in  
the ratio of electron-type to muon-type events of the different samples. 
Uncertainties in the ratio between different angular bins
are treated, similarly to Ref.~\cite{fogli2}. 
With our definitions we 
obtain, for instance,  $\chi^2_{SM}=$122/(35 d.o.f) which means that
the SM has a CL of $10^{-11}$ !

Next we minimize the $\chi^2$ function in Eq.~(\ref{chi2}) and determine the
allowed region in the parameter space for certain conversion
mechanism and  for a given confidence level, defined as, 
$\chi^2 \equiv \chi_{min}^2  + 4.61\ (9.21) $ for $90\ (99)$ \% C.L.

In Table \ref{tab:data} I show the minimum value of 
$\chi^2$ and the best fit point for several conversion mechanisms and for 
the different data sets. 
\begin{table*}[h]
\caption{Minimum value of $\chi^2$ and the best fit point for each 
channel and for different data sets. }
\label{tab:data}
\begin{center}
\begin{tabular}{|l|ll|l|l|l|}
\hline
Experiment & Oscillation  & $\nu_{\mu} \to  \nu_\tau$ &  
                $\nu_{\mu} \to  \nu_s$   &
                $\nu_{\mu} \to  \nu_s$   & 
                FC $\nu$-matter 
\\
           &   &  &      $\Delta m^2 < 0$ & $\Delta m^2 > 0$ &    \\ 
\hline
Super-Kam   & $\chi^2_{min}$ 
                           & $8.8$ & $12.9$  & $12.6$   & $9.4$    \\
contained                 & $ \Delta m^2 $ ( $10^{-3} $eV$^2$ )
             & $ 2.5 $    & $3.2$ &  $3.0$
& $\epsilon$=$0.95$  \\
  d.o.f=20-2 
                           & $\sin^2 2\theta$  
                           &  $0.98$   & $1.$   &  $0.99$ 
& $\epsilon'$=$0.084$  \\
\hline
Super-Kam   & $\chi^2_{min}$ 
                             & $1.3$  & $2.4$   & $2.3$  & $1$    \\
Stopping--$\mu$         & $ \Delta m^2 $ ( $10^{-3} $eV$^2$ )
                          & $ 3.0$    & $3.3$   & $3.7$
& $\epsilon$=$0.76$  \\
  d.o.f=5-2                            & $\sin^2 2\theta$ 
                           &  $0.99$   &  $1.$   &  $0.93$
& $\epsilon'$=$0.19$  \\
\hline
Super-Kam   & $\chi^2_{min}$ 
                             & $10.4$ &  $13.4$ & $10.5$  & $10.4$    \\
 Through-Going--$\mu$        & $ \Delta m^2 $ ( $10^{-3} $eV$^2$ )
            &$ 10.0 $ & $4.9$  & $18.$    
& $\epsilon$=$0.08$      \\
       d.o.f=10-2          & $\sin^2 2\theta$ 
                           &  $0.78$   & $1.$  & $0.55$  
& $\epsilon'$=$0.26$     \\
\hline
Super-Kam   & $\chi^2_{min}$ 
                             & $23.5$  & $32.9$ & $32.5$  & $43.8$   \\
  Combined                   & $ \Delta m^2 $ ( $10^{-3} $eV$^2$ )
                    & $3.1$      & $4.1$    &  $4.5$
& $\epsilon$ =$ 0.57$     \\
 d.o.f=35-2                & $\sin^2 2\theta$ 
                           &  $0.98$ &  $1.$  &   $0.96$
& $\epsilon'$=$0.45$      \\
\hline
\end{tabular}
\end{center}
\end{table*}

\begin{figure*}
\centerline{\protect\hbox{\epsfig{file=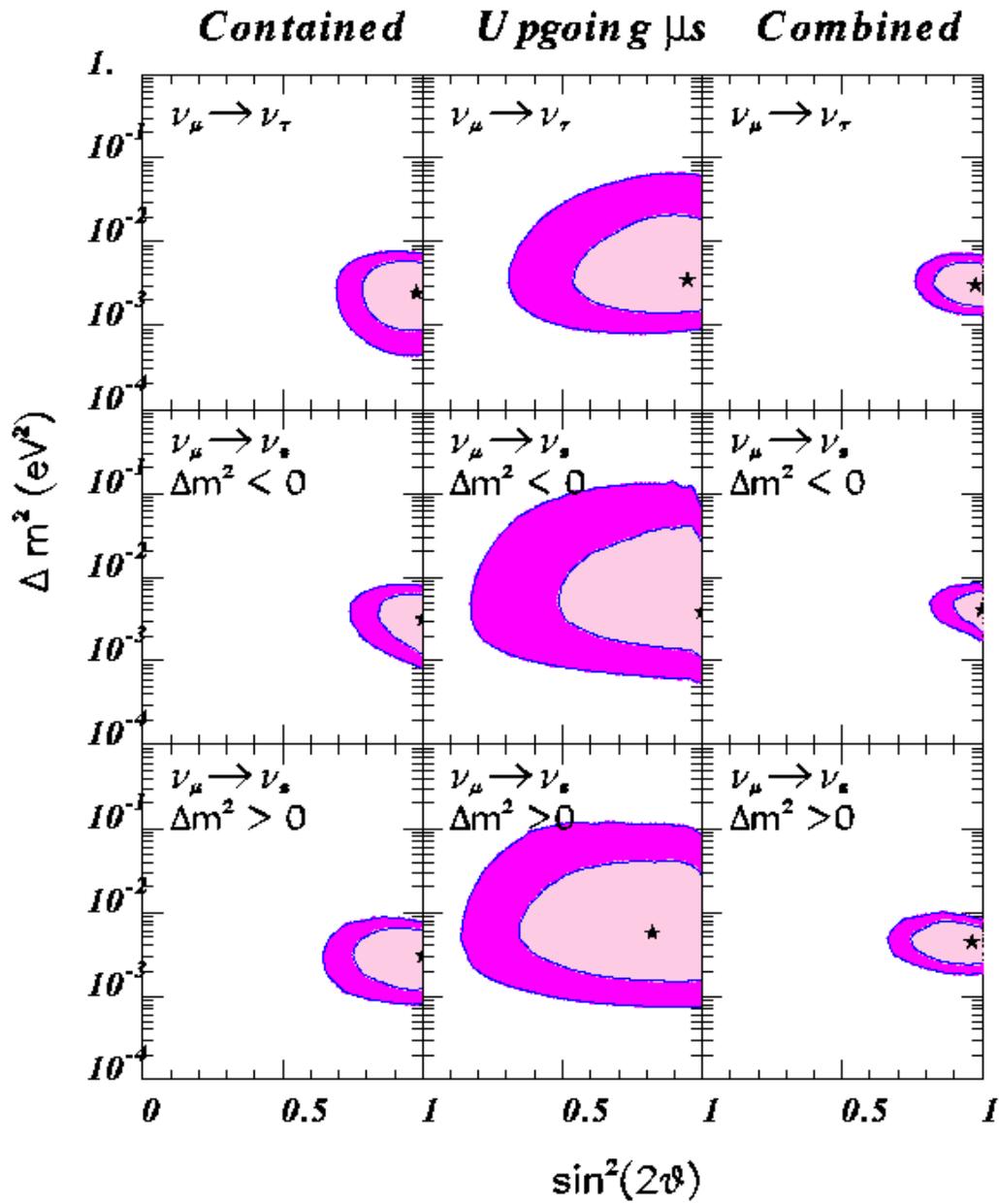,width=0.9\textwidth,
height=0.9\textheight}}}
\caption{
Allowed regions of the oscillation parameters for the different 
Super-Kamiokande data samples and oscillation channels as labelled
in the figure.} 
\label{cont_osc}
\end{figure*}

The results of our $\chi^2$ fit of the Super-Kamiokande contained and
upgoing atmospheric neutrino data in the framework of
neutrino oscillations are given in Fig.~(\ref{cont_osc}).
In this figure I give the allowed region of oscillation parameters at 90
and 99 \% CL.  

One can notice that matter effects lead to differences between the
allowed regions in the different channels. 
In the case of
$\nu_\mu \to \nu_s$ with $\Delta m^2>0$ matter effects enhance the 
oscillations for {\sl neutrinos} and therefore smaller values of the
vacuum mixing angle would lead to larger conversion probabilities
and the regions are therefore larger as compared to the vacuum case
$\nu_\mu\rightarrow\nu_\tau$. In the case of
$\nu_\mu \to \nu_s$ with $\Delta m^2<0$ 
the enhancement occurs only
for {\sl anti-neutrinos} while in this case the effect of matter
suppresses the conversion in $\nu_\mu$'s.  Since the yield of
atmospheric neutrinos is bigger than that of 
anti-neutrinos, clearly the matter effect suppresses the overall
conversion probability. Therefore one needs in this case a larger value
of the vacuum mixing angle, as can be seen by comparing the regions
in the second row with the corresponding ones in the first and third
row in Fig.~(\ref{cont_osc}).

Notice that in all channels where matter effects play a role 
the range of acceptable $\Delta m^2$ is slightly
shifted towards larger values, when compared with the $\nu_\mu \to
\nu_\tau$ case. This follows from looking at the relation between 
mixing {\sl in vacuo} and in matter. In fact, away from the
resonance region, independently of the sign of the matter potential,
there is a suppression of the mixing inside the Earth. As a result,
there is a lower cut in the allowed $\Delta m^2$ value, and it lies
higher than what is obtained in the data fit for the $\nu_\mu \to
\nu_\tau$ channel.  

\begin{figure}
\centerline{\protect\hbox{\epsfig{file=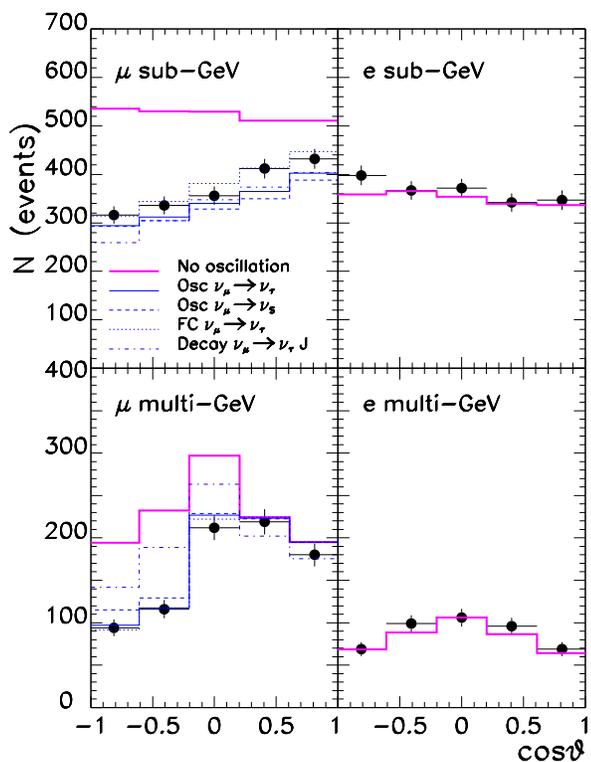,width=0.5\textwidth,height=0.5\textheight}}}
\caption{Angular distribution for Super-Kamiokande electron-like and muon-
like sub-GeV and multi-GeV events together with our prediction in the
absence of oscillation as well as the prediction for the
best fit point to the contained event data for the different conversion 
mechanism as labelled in the figure. The error displayed in the experimental 
points is only statistical.}
\label{ang_cont}  
\end{figure}

\begin{figure}
\centerline{\protect\hbox{\epsfig{file=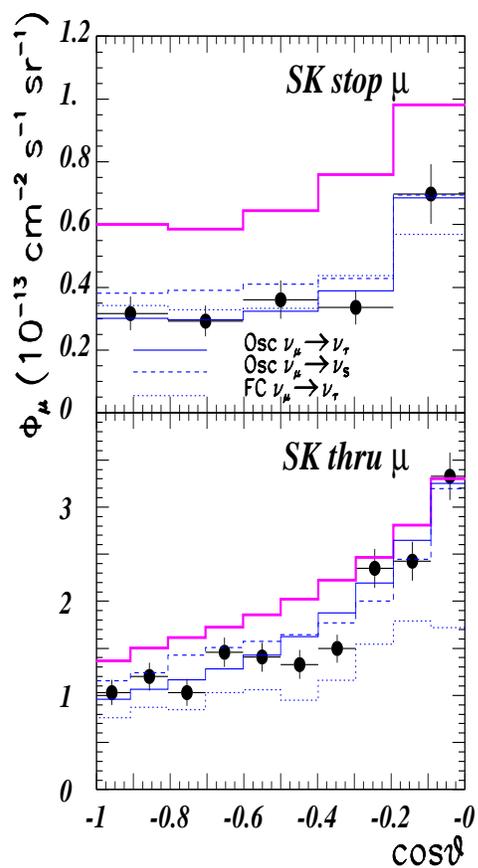,width=0.4\textwidth,
height=0.6\textheight}}}
\caption{Angular distribution for Super-Kamiokande upgoing muon data
together with our prediction in the
absence of oscillation  as well as the prediction for the
best fit point to the full data sample for the different conversion 
mechanism as labelled in  the figure. }
\label{ang_up}  
\end{figure}

Concerning the quality of the fits we see in table \ref{tab:data}
that the best fit to the full sample is obtained for the 
$\nu_\mu \to \nu_\tau$ channel although from the global analysis
oscillations into sterile neutrinos cannot be ruled out. 
One can also observe an improvement in the quality of the fits
to the contained events as compared to previous analysis performed with
lower statistics \cite{atm98}. These features can be easily 
understood by looking at the predicted zenith angle distribution of
the different event type for the various oscillation channels which 
I show in Fig.~(\ref{ang_cont}) for contained events and 
Fig.~(\ref{ang_up}) for upgoing muons. In Fig.~(\ref{ang_cont}) we observe
a perfect agreement between the observed distributions  of e-like events
and the predictions in the SM. This has lead to an improvement of
the quality of the description for any conversion mechanism that only 
involves muons. Also in Fig.~(\ref{ang_up}) we can observe that due
to matter effects the distribution for upgoing muons in the case of 
$\nu_\mu \to \nu_s$ are flatter than for $\nu_\mu \to \nu_\tau$ 
\cite{lipari,lipari1}.
Data show a somehow steeper angular dependence which can be better
described by $\nu_\mu \to \nu_\tau$. This leads to the better quality
of the global fit in this channel. Pushing further this feature 
Super-Kamiokande  collaboration has presented a preliminary partial 
analysis of the angular  dependence of the through-going muon data in 
combination with the up-down asymmetry of partially contained events 
which seems to exclude the possibility $\nu_\mu \to \nu_s$ at 
the 2--$\sigma$ level \cite{sk99}. 

\section{Alternative Scenarios}
As we have seen in the previous section the oscillation hypothesis 
provides a very good  explanation to the atmospheric neutrino data, and 
it is also simple and well-motivated theoretically. However,
alternative explanations to the atmospheric neutrino data have been
proposed in the literature. In this section I concentrate on the
present status of two possible "exotic''scenarios:  
FC-neutrino interactions in the Earth matter \cite{fcnc} and 
neutrino decay \cite{decay}.
\begin{figure}
\centerline{\protect\hbox{\epsfig{file=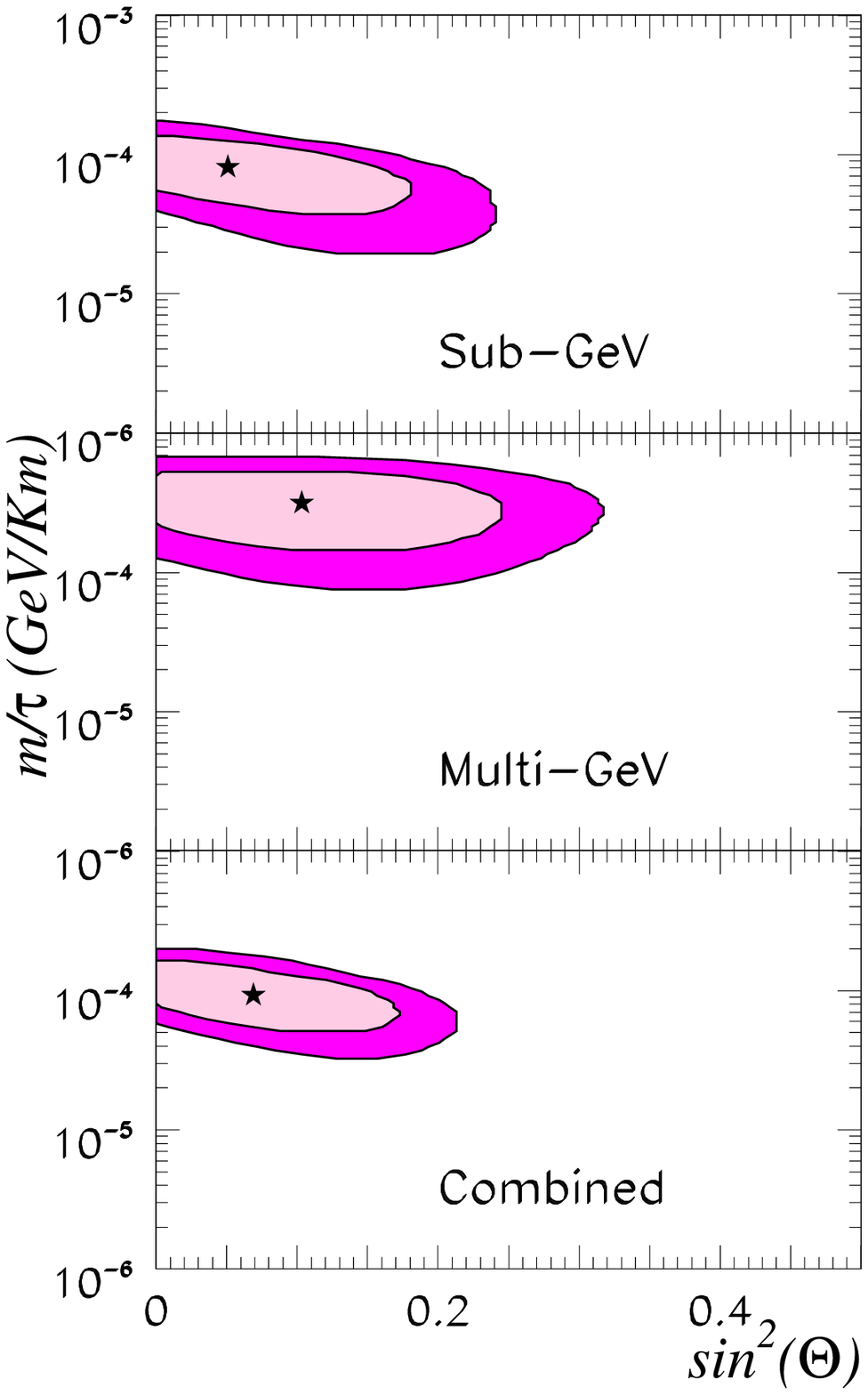,width=0.4\textwidth,
height=0.5\textheight}}}
\caption{
Allowed regions of th neutrino decay parameters for the different 
Super-Kamiokande data samples.} 
\label{contdecay}
\end{figure}

In Fig.~(\ref{contdecay}) I show the allowed regions in the parameter
space for the decay mechanism $\nu_\mu\rightarrow \nu_\tau  X$
for Super-Kamiokande sub-GeV and multi-GeV events separately as
well as the combined contained events. Partial fits to the 
two events samples lead to rather good description as can be seen
in Table \ref{tab:dec}.
\begin{table}[h]
\caption{Minimum value of $\chi^2$ and the best fit point to contained
events for the decay $\nu_\mu \to  \nu_\tau X$.}
\label{tab:dec}
\begin{center}
\begin{tabular}{|l|l|l|}
\hline
Experiment &   &  \\
\hline
Super-Kam   & $\chi^2_{min}$ &  1.4                    \\
 sub-GeV    & $m/\tau$ ($10^{-5}$ GeV/Km)&   8.1       \\
            & $\sin^2\theta$           &    0.05        \\
\hline
Super-Kam   & $\chi^2_{min}$ & 5.9                     \\
 multi-GeV  & $m/\tau$ ($10^{-5}$ GeV/Km)&  31.        \\
            & $\sin^2\theta$           &   0.1         \\
\hline
Super-Kam   & $\chi^2_{min}$ &     21.9            \\
Contained   & $m/\tau$ ($10^{-5}$ GeV/Km)&  9.3    \\
            & $\sin^2\theta$       &        0.07   \\
\hline
\end{tabular}
\end{center}
\end{table}
However the description of the global contained event sample is considerably
worse than in the case of oscillations. This arises from the stronger energy
dependence of the survival probability while the contained data
both in the sub-GeV and multi-GeV samples present a similar deficit.
As a consequence the allowed decay lifetimes which give a good description
to the  sub-GeV and multi-GeV data present very little overlap as can be
observed by comparing the upper and central panels in Fig.~(\ref{contdecay}).
As a consequence the mechanism gives a worse fit to the global contained
sample. This can also be observed in the angular distribution of contained
events for the best fit to the contained events (dash-dotted line
in Fig.~(\ref{ang_cont})) from where one sees that the decay hypothesis
cannot produce enough up-down asymmetry for the multi-GeV sample
without conflicting with the sub-GeV data.

This behaviour becomes particularly lethal when trying to 
describe the upward going muon data since for lifetimes favoured
by the contained event data very little muon conversion is expected
already for stopping muons in contradiction with observation. 
Based on this fact this mechanism
was ruled out in its simpler form in Ref. \cite{lipari,lisiup}
Recently the possibility of neutrino decay in a more general four neutrino 
scenario has been revisited in Ref.\cite{decaynew} where it is discussed
that a good description to the full atmospheric data sample is possible.

\begin{figure}
\centerline{\protect\hbox{\epsfig{file=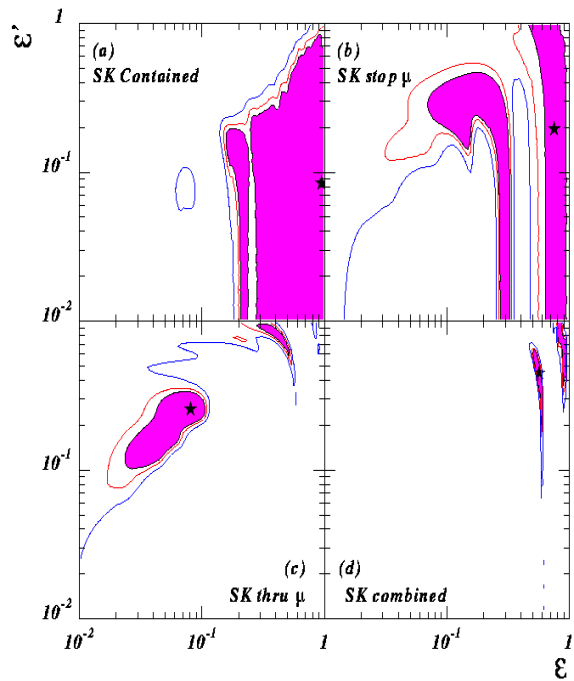,width=0.5\textwidth,
height=0.5\textheight}}}
\caption{
Allowed regions of the FC-neutrino interactions parameters for the different 
Super-Kamiokande data samples.} 
\label{contfcnc}
\end{figure}

I finally discuss our results on the alternative explanation of the 
atmospheric neutrino data in terms of FC neutrino-matter interactions 
\cite{fcnc}.
In Fig.~(\ref{contfcnc}) I show the contours of the regions allowed by
the Super-Kamiokande data. The different panels of the figure refer to
the fits performed over the different sets of data. 
The shaded areas are the regions allowed at 90\% C.L., while
the dashed and dotted contours refer to 95 and 99 \% C.L.,
respectively. In Table \ref{tab:data} I list the corresponding
best fit points as well as the values of $\chi^2_{min}$ attainable
for this mechanism. From the table we see that contained and 
stopping muon events can be described with a quality comparable
to the oscillation channels. In Fig.(\ref{ang_cont}) one can also
observed that the FC-neutrino mechanism leads to a very good
description of the zenith angular distributions for contained
events. This may come as a surprise since angular distributions
for multi-GeV and sub-GeV events are rather different while
the FC mechanism leads to an energy independent conversion 
probability. One must bear in mind, however, that the plotted 
angular distribution is that of the produced charged leptons
in the neutrino interaction. For neutrinos leading to sub-GeV
events the average opening angle between the neutrino and the
produced lepton is $60$ degrees what leads to the flat observed 
distribution almost independently of the specific conversion mechanism.
The overall normalization, is however, totally consistent with the
FC hypothesis as the deficit in both samples is of about 60 \%.

The allowed regions
can be qualitatively understood in the approximation of constant
matter density in Eq.(\ref{probfc}). From there one can see that in order 
to have a  relatively large transition probability, as implied by the
contained events and also by the stopping muons events,
the FC parameters are required to be in the region 
$\epsilon^{\prime}_\nu < \epsilon_\nu$ and $\eta> \pi/R_{\oplus}$. This last
condition leads to a lower bound on $\epsilon_\nu$.  The island in
Fig.~(\ref{contfcnc}.a) corresponds to $\eta\sim\pi/R_{\oplus}$. 
On the other hand, the through-going muon data require a smaller transition 
probability and therefore the region 
$\epsilon^{\prime}_\nu > \epsilon_\nu$ turns out to
be the preferred one. 

The combination of the different data sets in a single
$\chi^2$-analysis is shown in Fig.~(\ref{contfcnc}.d).  
As seen in the figure as well as in Table \ref{tab:data}, 
when the angular information  of both
stopping and through-going muons is included in the data analysis, 
the quality of the full description  worsens and leads to 
a $\chi^2_{min}=43.8/(33 d.o.f)$. Since FC-neutrino interactions
lead to an energy independent conversion probability, the smaller
deficit in the through-going muon sample as compared to contained
and stopping-muon samples cannot be well accommodated. 
As seen in the lower panel in Fig.~(\ref{ang_up}) the prediction for 
the best fit point for this  mechanism would imply a larger deficit
for higher energy muons mainly in  the last three angular 
bins where it does not produce a sufficient amount of through-going muons
at angles  $ 0< \theta <20 $ degrees below the horizon.

\end{document}